\documentclass[twocolumn,showpacs,floatfix,superscriptaddress, showkeys,prc]{revtex4-2}
\usepackage{mathrsfs}
\usepackage{graphicx}
\usepackage{dcolumn} 
\usepackage{bm} 
\usepackage[colorlinks,linkcolor=blue,citecolor=blue]{hyperref}
\usepackage{amsmath, amssymb}
\usepackage{multirow}
\usepackage{CJK}
\usepackage[section]{placeins}

\newcommand{\svec}[1]{\boldsymbol{#1}}

\newcommand{\cals}[1]{{\mathcal #1}}

\newcommand{\lrlc}[1]{\left|#1\right>}
\newcommand{\lrcl}[1]{\left<#1\right|}
\newcommand{\lrs}[1]{\left[#1\right]}

\newcommand{\sigs}{{\sigma\text{-S} }}

\newcommand{\add}[1]{\textcolor[rgb]{1.00,0.00,0.00}{#1}}

\newcommand{\ti}{{\textsf{i}}}

\newcommand{\td}{{\textsf{d}}}

\newcommand{\tD}{{\textsf{D}}}
\newcommand{\tE}{{\textsf{E}}}

\newcommand{\tIm}{\operatorname{Im}}

\begin{document}

\title{Single-Particle Resonant States in Relativistic Hartree-Fock Theory: A Green's Function Approach }
 \author{Wei Gao }
 \affiliation{Frontier Science Center for Rare isotope, Lanzhou University, Lanzhou 730000, China}
 \author{Ting Ting Sun }\email{ttsunphy@zzu.edu.cn} 
 \affiliation{School of Physics, Zhengzhou University, Zhengzhou 450001, China}
 \author{Wen Hui Long }\email{longwh@lzu.edu.cn} 
 \affiliation{Frontier Science Center for Rare isotope, Lanzhou University, Lanzhou 730000, China}
 \affiliation{School of Nuclear Science and Technology, Lanzhou University, Lanzhou 730000, China}
 \affiliation{Joint Department for Nuclear Physics, Lanzhou University and Institute of Modern Physics, CAS, Lanzhou 730000, China}
\begin{abstract}
Relativistic Hartree-Fock theory is combined with the Green's function method in coordinate space to study both single-particle bound and resonant states within a unified framework. Within this approach, single-particle resonance energies and widths are unambiguously extracted from the density of states, and the influence of the Coulomb exchange effects on proton resonances in $N=82$ isotones are systematically examined. It is found that the exact treatment of  Coulomb exchange terms reduces proton resonance energies of approximate $0.09\sim0.21$ MeV,  a significantly smaller effect than that obtained from the phenomenological treatment. Moreover, except for rather narrow resonances, the proton resonance widths are visibly reduced by the Coulomb exchange terms, also being much less pronounced than the phenomenological approach. Notably, clear shell effects are observed in the isotonic evolutions of the resonance energy reductions for specific resonances. All these highlight the necessity of a microscopical and exact treatment of the Coulomb exchange terms. 
\end{abstract}
\maketitle 

\section{Introduction}
 
Resonance states are essential for understanding the exotic properties in nuclear structure, nuclear reaction and scattering processes, as well as nuclear astrophysics. In particular, the continuum contributions dominated by resonant states play an essential role in the formation of the novel nuclear halo structures \cite{Tanihata1985PRL55.2676, Meng1996PRL77.3963, ZhangSS2013EPJA49.77, GuoJY2019PRC99.024314, Demyanova2020PRC102.054612, Meng1998PRL80.460, ZhangY2012PRC86.12, ZhangSS2012EPJA48.40, Sandulescu2003PRC68.054323, Zhou2010PRC82.011301, Li2012PRC85.024312, GuoJY2019PRC99.014309, Watanabe2024PRC110.054608}. Similarly, it has been found that the continuum contributions to the collective giant resonances are also dominated by the single-particle resonances \cite{CaoLg2002PRC66.024311, Daoutidis2009PRC80.024309}. Moreover, the descriptions of the mirror symmetry breaking in the O isotopes \cite{ZhangS2022PLB827.136958} and the low-lying excitation spectra of the C isotopes \cite{GengYF2022.PRC106.024304} can be improved by considering the degrees of freedom associated with d{resonant states. In fact, the energy and width of resonant states serve as essential inputs for studying stellar nucleosynthesis \cite{ZhangSS2012PRC86.032802, Faestermann2015PRC92.052802}.

To date, the methodology based on the scattering theory has been well established for studying resonant states. As an effective strategy, the energies and widths of resonant states can be determined by analyzing the $R$-matrix and exploiting its connection to the collision matrix \cite{Wigner1947PR72.29, Hale1987PRL59.763, Descouvemont2010RPP73.036301}. Combined with the Breit--Wigner formula or the full width at half maximum of the scattering cross section, the $S$-matrix method \cite{book.Taylor1972, CaoLg2002PRC66.024311} and scattering phase shift method \cite{Ferreira1997PRL78.1640, Sandulescu2003PRC68.054323, Hamamoto2005PRC72.024301, Fortune2006PRC73.014318, LiZp2010PRC81.034311} ascertain the energies and widths of resonant states by analyzing scattering phase shifts, which provide a more explicit and intimate correspondence with experimentally measured reaction cross sections. As a precise and well-established alternative, both bound states and resonant states can be determined by locating the zeros of the Jost function \cite{Sofianos1999PRAv60.337, LvBN2012PRL109.072501, LvBN2013PRC88.024323}. Nevertheless, the numerical implementation of the Jost function approach remains challenging.

For the sake of simplicity, the methods used for bound states can also be extended to identify single-particle resonant states. For instance, resonance energies and widths can be extracted by examining the energy dependence on the space size adopted in solving the Schr\"odinger or Dirac equations, giving rise to the real stabilization method (RSM) \cite{Hazi1970PRA1.1109_RSM, Mandelshtam1993PRL70.1932_RSM, ZhangL2007APS56.3839_RSM, ZhangL2008PRC77.014312_RSM, Lay2012PRC85.054618, YangW2024APS73.062102}. Alternatively, by transforming resonant states into bound ones, e.g., with artificially enhanced attractive potential, the energies, widths, and wave functions of resonant states can be obtained via a Pad\'e approximant, namely the Analytic Continuation of the Coupling Constant (ACCC) method \cite{book.Kukulin1989, Cattapan2000PRC61.067301, YangSc2001CPL18.196, ZhangSS2004PRC70.034308, GuoJY2005PRC72.054319, DongXX2015PRC92.024324}. As another alternative strategy, the complex scaling method (CSM) \cite{Gyarmati1986PRC34.95, Moiseyev1998PR302.212, GuoJY2010PRC82.034318, LiuQ2012PRC86.054312, GuoJY2014PRC90.034319, ShiM2015PRC92.054313}, which rotates the single-particle Hamiltonian in the complex coordinate space, has been also developed to explore resonant states. Furthermore, by transforming into the complex momentum space, the complex momentum representation (CMR) method \cite{LiN2016PRL117.062502, Tian2017CPC41.044104, ShiM2018PRC97.064301} has been derived to precisely determine resonant states, thereby circumventing the need to handle non-square-integrable wave functions in coordinate space. 

Green's function (GF) method is another effective approach to explore the single-particle resonant states. By adopting  the boundary condition derived from the scattering theory, this method enables an appropriate description of the asymptotic behavior of continuum wave functions \cite{book.Economou2006green, Tamura1992PRB45.3271, Shlomo1975NPA243.507518, Belyaev1987SJNP45.1263, Wehrberger1988PRC37.1148, Shepard1989PRC40.2320}. In particular, given its poles in the complex-energy plane, both bound and resonant states can be determined precisely, and the single-particle density distribution can be conveniently obtained via contour integration around the poles. Owing to such features, the GF method provides a unified framework for treating bound states and the continuous spectrum. It not only allows for the accurate determination of energies and widths of arbitrary-width resonances in the continuum \cite{SunTT2020CPC44.084105,SunTT2021NST32.46}, but also properly describes the asymptotic behavior of spatial density distributions, which lays a foundation to describe nuclear collective excitations, as combined with various theoretical models. 

Practically, the GF method has been widely applied in nuclear structure research. {By considering the pairing correlations, it has been extended to explore the influence of the continuum on the ground-state properties of spherical nuclei in the drip line region \cite{ZhangY2011PRC83.054301, ZhangY2012PRC86.054318, ZhangY2014PRC90.034313, SunTT2019PRC99.054316, SunTT2023NST34.105, SunTT2016SSPMA46.012006}, and further for deformed nuclei \cite{Oba2009PRC80.024301, SunTT2020PRC101.014321}. Moreover, the GF method based on the random phase approximation has been established to investigate the contribution of the continuum to collective excitations of nuclei \cite{Matsuo2001NPA696.371395, CaoLg2002PRC66.024311, Daoutidis2009PRC80.024309, YangD2010PRC82.054305, YangD2010CTP53.716722, YangD2010CTP53.723730, Daoutidis2011PRC83.044303, Sun2024EPJA60.61}. Within the relativistic scheme, the GF method has been also developed to extract resonant states \cite{SunTT2014PRC90.054321}, providing a unified description of the pseudo-spin symmetry in both bound and resonant states \cite{SunTT2019PRC99.034310, SunTT2023PLB847.138320, SunTT2024PLB854.138721}. Particularly, the evolution law of spin and pseudo-spin symmetries in deformed nuclei and hypernuclei has been profoundly revealed \cite{SunTT2017PRC96.044312, SunTT2017JPG44.125104, SunTT2024PRC109.014323}.

In the past decades, the relativistic mean-field (RMF) theory has achieved great success in nuclear structure research, and has become one of the most important microscopic approaches for systematically describing the structural properties of atomic nuclei \cite{Serot1986ANP16.1,Ring1996PPNP37.193,Vretenar2005PRe409.101,Niksic2011PPNP66.519}.
However, the exchange Fock terms are dropped in the RMF models only for the sake of simplicity. After implementing the Fock terms, the relativistic Hartree-Fock (RHF) theory \cite{Bouyssy1987PRC36.380, Long2006PLB640.150, Long2010PRC81.024308, GengJ2020PRC101.064302, GengJ2022PRC.105.034329, PengY2025CPC49.074107_ORHF}, with the proposed RHF Lagrangians PKO$i$ ($i = 1,2,3$) \cite{Long2006PLB640.150, Long2008EPL82.12001} and PKA1 \cite{Long2007PRC76.034314}, has achieved comparable quantitative precision on the description of nuclear structural properties as popular RMF models. In particular, the degrees of freedom associated with the $\pi$- and $\rho$-tensor couplings, which contribute almost fully via the Fock terms, present strong impact on the binding of nuclear systems \cite{Geng2019PRC100.051301R}. This has brought about significant improvements in the self-consistent description of spin-isospin excitations \cite{Liang2008PRL101.122502, Liang2012PRC85.064302, Niu2013PLB723.172, Niu2017PRC95.044301}, new magicity \cite{Li2016PLB753.97, Li2019PLB788.192}, and novel nuclear phenomena \cite{Li2019PLB788.192, GengJ2023CPC47.044102, GengJ2024PLB858.139036, PengY2025CPC49.064112}. Hence, it is worthy to elucidate the effects of the Fock terms on single-particle resonant states based on the RHF theory.

Proton resonant states are of profound significance for understanding proton halos \cite{ZhangSS2013EPJA49.77, Demyanova2020PRC102.054612}, and proton capture and emission processes \cite{Chong2009SCSG, Nguyen2022PRC106.L051302, GuoJY2024PRC110.014323}, in which the Coulomb interactions are of fundamental significance. Particularly, it has been found that the isospin symmetry-breaking corrections to super-allowed $\beta$ decays, which are crucial for testing the unitarity of the Cabibbo-Kobayashi-Maskawa matrix, are sensitive to the Coulomb exchange terms \cite{Liang2009PRC79.064316, Towner2010PPP73.046301}. In fact, applying the CSM, the influence of the Coulomb exchange terms on resonant states has been analyzed in a phenomenological way \cite{Niu2013PRC87.037301, GuoJY2014PRC89.034307, NiuZM2016NST27.122}. It is worth noting that, the Coulomb exchange terms are treated exactly within the RHF theory, and thus the relevant impact on proton resonant states is warranted to be analyzed in detail.

Given the advantages in addressing resonant states \cite{Daoutidis2009PRC80.024309, YangD2010PRC82.054305, YangD2010CTP53.716722, YangD2010CTP53.723730, Daoutidis2011PRC83.044303, SunTT2014PRC90.054321, SunTT2016SSPMA46.012006, SunTT2023PLB847.138320, SunTT2024PLB854.138721, Sun2024EPJA60.61, SunTT2024PRC109.014323}, the GF method based on the RHF theory, namely the RHF-GF method, is proposed in this work, with a special focus on the influence of the exact Coulomb exchange terms in proton resonances. The general formalism of the RHF-GF method is introduced briefly in Section \ref{sec:general formalism}. Section \ref{sec:result and discussions} presents the discussions of the results, encompassing the validation of the RHF-GF method by taking $^{120}$Sn as a benchmark, and systematic investigation of proton resonant states in $N=82$ isotones and the underlying Coulomb exchange effects. Finally, a concise summary is provided in Section \ref{sec:summary}.

\section{The RHF-GF Method}\label{sec:general formalism}
Within the RHF framework, the exchange correlations are incorporated by introducing the explicit Fock terms. In contrast to the local Hartree terms, the Fock terms inherently couple the wave function at different positions $\svec x$ and $\svec x'$, which leads to the non-local potential \cite{Bouyssy1987PRC36.380, Long2006PLB640.150, Long2007PRC76.034314, Long2008EPL82.12001, Long2010PRC81.024308, GengJ2020PRC101.064302, GengJ2022PRC.105.034329, PengY2025CPC49.074107_ORHF}. This results in an integro-differential Dirac equation for the single-particle motion inside nucleus, which is hard to solve directly in coordinate space. Practically, such equation is solved by introducing equivalent localization of the non-local potential \cite{Bouyssy1987PRC36.380, Long2006PLB640.150}, or utilizing the basis expansion method \cite{Long2010PRC81.024308, GengJ2020PRC101.064302, GengJ2022PRC.105.034329, PengY2025CPC49.074107_ORHF}. However, these methods are generally limited to bound states or discrete continuum states. In this work, GF method is applied to solve the spherical RHF equation in the coordinate space, namely the radial Dirac equation, by which the bound and resonant states in continuum are determined on the same footing. 

\subsection{Green's function of integro-differential Dirac equations}
Under the relativistic scheme, the spherical single-particle wave function, namely the Dirac spinor, can be expressed 
as,
\begin{align}\label{eq:dirac_spinor}
\psi_{\alpha}(\svec x)=\frac{1}{r}\begin{pmatrix} G_a^+(r) \Omega_{+\kappa m}(\vartheta, \varphi)\\[0.5em] \ti G_a^-(r) \Omega_{-\kappa m}(\vartheta, \varphi) \end{pmatrix}\chi_{\tau_\alpha},
\end{align}
where $\svec x = r \svec e_r(\vartheta,\varphi)$, $G_a^\pm$ represent the radial wave functions for the upper ($+$) and lower ($-$) components, $\Omega_{\kappa m}$ (also referred as $\Omega^l_{j m}$ ) is the spherical spinor, and $\chi_{\tau_\alpha}$ denotes the isospin wave function with isospin projection $\tau_\alpha$. Here, the Greek subindex $\alpha = (am)$ denote the single-particle states, with $a = (n\kappa)$ and $\kappa = \pm (j+1/2)$ for $j = l\mp1/2$, and $n$, $j$ ($m$) and $l$ for the principal quantum number, the total angular momentum (projection) and the orbital one, respectively. 

From the variation of the RHF energy functional \cite{Bouyssy1987PRC36.380, Long2010PRC81.024308}, the radial Dirac equation can be formally derived as,
\begin{align}\label{eq:RHF}
  \int \hat{h}(r,r')\Psi_a(r') \td r'= & \varepsilon_{a}\Psi_{a}(r), & \Psi_{a} \equiv & \begin{pmatrix}G_a^+ \\[0.5em]    G_a^- \end{pmatrix},
\end{align}
where $\varepsilon_{a}$ represents the single-particle energy excluding the rest mass $M$, and $\Psi_a$ represents the combination of radial wave functions. For the single-particle Dirac Hamiltonian $\hat{h}(r,r')$, it contains the kinetic energy term $h^{\text {kin}}(r,r')$, the local term $h^\tD(r,r')$, and the non-local potential $h^\tE(r,r')$,
\begin{subequations}\label{eq:RHF-H}
  \begin{align}
h^{\text {kin}} & =[\svec\alpha \cdot \svec p +(\beta -\mathbb{E}) M] \delta(r-r'), \\
h^\tD & =\left[\Sigma_T(r) \gamma_5+\Sigma_0(r)+\beta \Sigma_S(r)\right] \delta(r-r'), \\
h^\tE & = \begin{pmatrix} \Sigma^{++}(r, r') & \Sigma^{+-}(r, r') \\[0.5em] \Sigma^{-+}(r, r') & \Sigma^{-}(r, r')\end{pmatrix},
\end{align}
\end{subequations}
where $\mathbb{E}$ represents the identity matrix.

In the single-particle Hamiltonian (\ref{eq:RHF-H}), the local self-energies $\Sigma_S$, $\Sigma_0$, $\Sigma_T$ are contributed by the direct (Hartree) terms, with $\Sigma_0$ containing the rearrangement terms originating from the density dependencies of the meson-nucleon coupling strengths \cite{Long2006PLB640.150, Long2007PRC76.034314, Long2010PRC81.024308, GengJ2020PRC101.064302}. The non-local potentials $\Sigma^{\mu\mu'}$ are contributed by the exchange (Fock) terms, with $\mu,\mu'=\pm$ denoting the upper and lower components of $\Psi_a$, respectively. Taking the $\sigma$-scalar ($\sigs$) coupling as an example, the non-local potentials can be uniformly expressed as,
\begin{align}
\Sigma_{\sigs, a}^{\mu \mu'}(r,r') = &g_\sigma(r) \sum_b \delta_{\tau_a\tau_b} v_b^2 (2j_b+1)\nonumber\\
&\hspace{2em}\times g_\sigma(r')\cals R_{\sigma;a b}^{\mu \mu'}(r, r') R^{\mu \mu'}_b(r, r'),
\end{align}
where $g_\sigma$ denotes the $\sigma$-S coupling strength, $v_b^2$ is the occupation probability of orbit $b$, and the symbol $\cals R_{\sigma;a b}^{\mu\mu'}$ represents the combination of the CG coefficients and the propagator term \cite{Long2010PRC81.024308}. In order to abbreviate expressions, the non-local density $ R^{\mu \mu'}_a(r,r')$ is introduced as,
\begin{align}\label{eq:sq_non-local_dens}
   R^{\mu \mu'}_a(r, r')=G_a^\mu(r)G_a^{\mu'*}(r').
\end{align}
For further details, please refer to Refs. \cite{Bouyssy1987PRC36.380, Long2010PRC81.024308, GengJ2020PRC101.064302}. It is obvious that if only considering the Hartree terms, the radial Dirac equation (\ref{eq:RHF}) is reduced to a differential equation as in the RMF theory. 

Consistent with the radial Dirac equation (\ref{eq:RHF}), the equation of the GF can be deduced as,
\begin{equation}\label{eq:RHF-GF-eq}
\int\lrs{\varepsilon\delta(r-r')-\hat{h}(r, r')}\cals G(r', r'';\varepsilon)\td r'=\mathbb{E} \delta(r-r''),
\end{equation}
where $\cals G( r' , r'' ; \varepsilon)$ is the GF with complex probing energy $\varepsilon$. Analytically, the operator of the GF can be expressed as the inverse of $\varepsilon-\hat h$, namely $\hat{\cals G}=(\varepsilon-\hat{h})^{-1}$. Applying the condition satisfied by the solutions of the radial Dirac equation (\ref{eq:RHF}), namely $\sum_n\lrlc{a}\lrcl{a}=1$, the operator $\hat{\cals G}_\kappa$ for given $\kappa$ quantity is further deduced as,
\begin{align}
 \hat{\cals G}_\kappa(\varepsilon) = & \sum_n(\varepsilon-\hat{h})^{-1}\lrlc{a}\lrcl{a} = \sum_n\frac{\lrlc{a}\lrcl{a}}{\varepsilon-\varepsilon_{a}}.
\end{align} 
Consistently, the GF is derived as, 
\begin{align}\label{eq:GF_eigen}
  \cals G_\kappa(r,r';\varepsilon) = & \lrcl{r} \hat{\cals G}_\kappa(\varepsilon)\lrlc{r'} = \sum_n \frac{\Psi_a(r)\Psi_a^\dag(r')}{\varepsilon - \varepsilon_a},
\end{align}
where $\Psi_a(r) = \left< r|a\right>$. Obviously, both the bound and resonant states can be characterised by the poles of the GF in the complex energy plane. Notice that the sum over $n$ in the above expression includes both the discrete and continuous spectra. Thus, the GF contains the information of bound states and the continuum, as well as resonant states.

As deduced from the form of $\Psi_a$, it is readily apparent that the GF assumes a form of density matrix. With the poles denoted by $\varepsilon_a$, the GF in Eq. (\ref{eq:GF_eigen}) is meromorphic on the complex energy plane. Notably, owing to its inherent non-locality, the GF provides a natural framework for treating the non-local exchange terms. In terms of the non-local density (\ref{eq:sq_non-local_dens}), the components of the GF can be uniformly formulated as
\begin{align}
  \cals G_\kappa^{\mu\mu'}(r,r';\varepsilon)=\sum_n\frac{R_a^{\mu\mu'}(r,r')}{\varepsilon-\varepsilon_a}.
\end{align}
Applying the Cauchy's theorem, the non-local density $R_a^{\mu\mu'}(r,r')$ can be obtained from the contour integration of the GF in the complex energy plane,
\begin{align}\label{eq:GF_non-local_density}
   R_a^{\mu\mu'}(r,r')=\frac{1}{2\pi \ti}\oint_{C_a} \td \varepsilon ~ \cals G_\kappa^{\mu\mu'}(r,r';\varepsilon),
\end{align}
where an contour $C_a$ is chose to enclose only the state $a$. As displayed in Fig. \ref{fig:integral_diagram}, one can select the contour $C_a$ sequentially for each state $a$ to deduce the non-local density $ R_a^{\mu\mu'}(r,r')$.

\begin{figure}[htbp]
  \centering
  \includegraphics[width=0.48\textwidth]{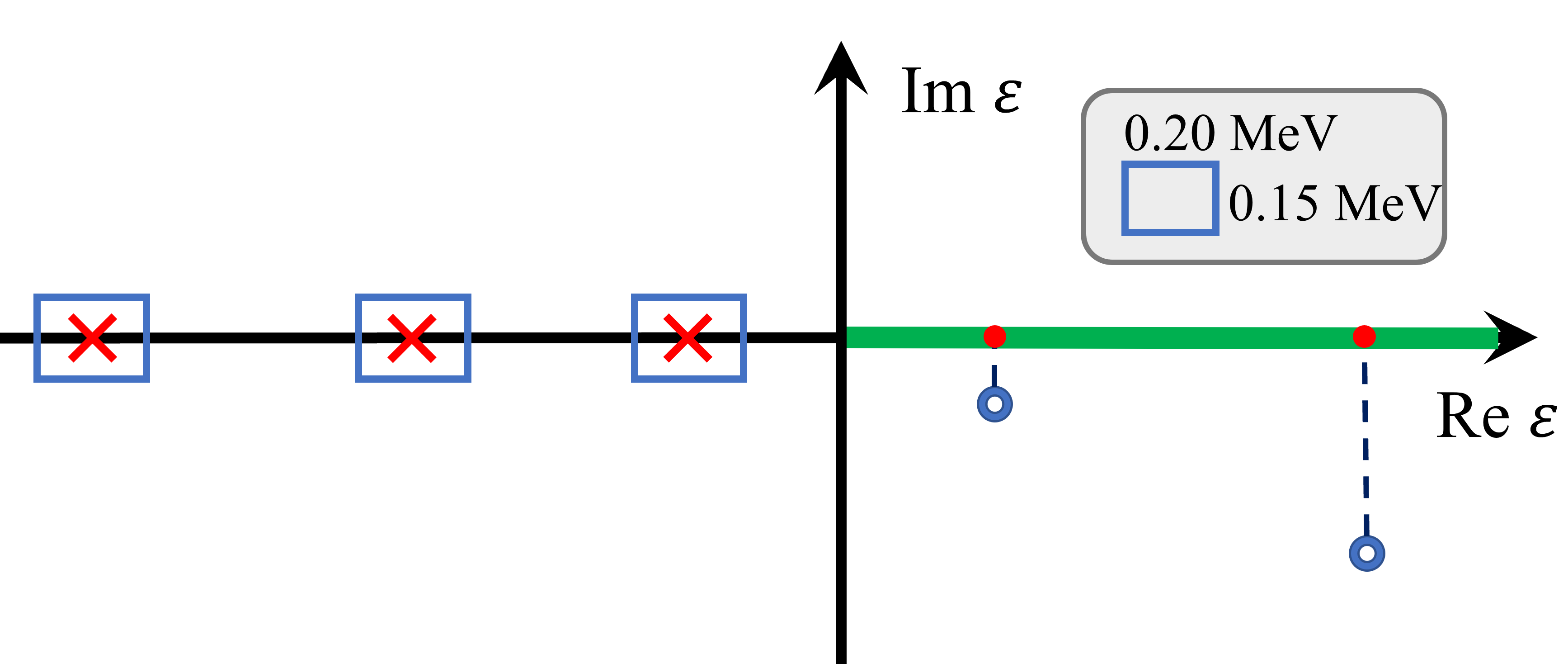}
  \caption{(Color online) Contours for the Green's function on the complex energy plane, with a rectangle measuring 0.15 MeV in width and 0.20 MeV in length, and the integration step size reading 0.01 MeV. The red crosses, blue hollow circles and thick green line represent bound states, resonant states $E - i\Gamma/2$, and the continuum states, respectively.}\label{fig:integral_diagram}
\end{figure}

In principle, the local density can be obtained from the non-local density $R_{a}^{\mu\mu'}(r,r')$ by setting  $r=r'$. Thus, from the contour integration of the GF, which gives the non-local density and local ones, one can calculate the local potential $h^{\tD}$ and the non-local potential $h^{\tE}$ in Eq. (\ref{eq:RHF}), which are further employed to construct a new GF that characterizes bound and resonant states. Iteratively, a self-consistent procedure is achieved within the RHF-GF framework.

\subsection{Construction of Green's Function}

However, it is not practically feasible to construct the GF from Eq. (\ref{eq:GF_eigen}), which only manifests its link to the bound and resonant states. To construct the GF , a piecewise method is universally adopted. Specifically, giving the probing energy $\varepsilon$, one may obtain two linearly independent column vectors from Eq. (\ref{eq:RHF}), namely the incoming wave function $\Phi_\kappa^{\text{in}}(r; \varepsilon)$ and the outgoing one $\Phi_\kappa^{\text{out}}(r; \varepsilon)$, via the Runge-Kutta method initiated from the asymptotic behaviors as $r \to \infty$ and $r \to 0$, respectively. It is expected that, as $r \to \infty$, wave functions oscillate for $\operatorname{Re}\varepsilon > 0$, and decay exponentially with $\operatorname{Re}\varepsilon < 0$. Thus, the boundary condition as $r\to \infty$, namely the asymptotic behavior derived from the scattering theory, is adopted to gain the incoming wave function $\Phi_\kappa^{\text{in}}(r; \varepsilon)$. For the outgoing one $\Phi_\kappa^\text{out}(r)$ starting from $r\to 0$, it is regular at the origin. For details, please refer to Refs. \cite{book.Economou2006green, SunTT2014PRC90.054321, SunTT2020PRC101.014321, YangD2010CTP53.716722}.

Given the probing energy $\varepsilon$ and quantum number $\kappa$, the GF $\cals G_{\kappa}(r,r';\varepsilon)$ can be obtained as \cite{Tamura1992PRB45.3271, Daoutidis2009PRC80.024309, SunTT2014PRC90.054321}
\begin{align}\label{eq:GF_construct}
\cals G _\kappa(r,  r' ; \varepsilon)&=\frac{1}{W_\kappa(\varepsilon)}\left[\theta(r-r') |\Phi_\kappa^{\text{in}}(r; \varepsilon) \rangle\langle \Phi_\kappa^{\text{out}*}(r'; \varepsilon)|\right.\notag\\[0.4em]
      &\left.+\theta(r'-r) |\Phi_\kappa^{\text{out}}(r; \varepsilon) \rangle\langle \Phi_\kappa^{\text{in}*}(r'; \varepsilon)|\right],
\end{align}
where $\theta(r-r')$ is the step function, and $\Phi_\kappa^{\text{out}}(r; \varepsilon)$ and $\Phi_\kappa^{\text{in}}(r; \varepsilon)$ correspond to the following combination,
\begin{align}
  \Phi_\kappa^{\text{out}}=\begin{pmatrix}g_\kappa^{+,\text{out}}(r; \varepsilon)\\[0.5em]g_\kappa^{-,\text{out}}(r; \varepsilon)\end{pmatrix}, \quad
  \Phi_\kappa^{\text{in}}=\begin{pmatrix}g_\kappa^{+,\text{in}}(r; \varepsilon)\\[0.5em]g_\kappa^{-,\text{in}}(r; \varepsilon)\end{pmatrix}.
\end{align} 
In Eq. (\ref{eq:GF_construct}), the Wronskian function $W_\kappa(\varepsilon)$ is defined as,
\begin{align}
  W_\kappa(\varepsilon)\equiv & \begin{vmatrix}
    g_\kappa^{+,\text{out}}(r; \varepsilon) & g_\kappa^{+,\text{in}}(r; \varepsilon) \\[0.5em]
    g_\kappa^{-,\text{out}}(r; \varepsilon) & g_\kappa^{-,\text{in}}(r; \varepsilon)
  \end{vmatrix}.
\end{align}
It is worth noting that, the function $W_\kappa(\varepsilon)$ is $r$-independent, given the Hermite single-particle Hamiltonian $\hat h$. Moreover, $W_\kappa(\varepsilon)$ replicates the denominator in Eq. (\ref{eq:GF_eigen}), namely $W_\kappa(\varepsilon) \to 0$ when the probing energy $\varepsilon$ approaches the pole $\varepsilon_a$.

Due to the Fock terms, the radial Dirac equation (\ref{eq:RHF}) is an integro-differential equation. To obtain the incoming and outgoing wave functions, the non-local terms in Eq. (\ref{eq:RHF}) shall be localized equivalently. In terms of the non-local density (\ref{eq:GF_non-local_density}), the equivalent local potential $V_a^{\mu\mu'}$ for orbit $a$ can be obtained as \cite{Bouyssy1987PRC36.380, Long2006PLB639.242},
\begin{align}
  V_a^{\mu\mu'}(r) = & \int\td r' \sum_\nu^{\pm} \frac{\Sigma_a^{\mu\nu}(r,r') R_a^{\mu'\nu} (r,r')}{ R_a^{++}(r,r) + R_a^{-}(r,r)},
\end{align}
where $\mu, \mu', \nu=\pm$ represents the upper ($+$) and lower ($-$) components of $\Psi_a$. It is worthy to comment that an inner iteration is considered to promise the convergence of the energies and widths of resonant states, since the localization depends on the non-local density derived from the GF.

\subsection{Density of states for resonance energy and width}

Bound and resonant states can be directly determined by locating poles of the GF in the complex-energy plane, and the density of states (DoS) enables us to rapidly and precisely identify these poles. As rigorously established in Refs.  \cite{book.Economou2006green,SunTT2014PRC90.054321, SunTT2020CPC44.084105}, the DoS $n(\varepsilon)$ can be derived from the integral of the imaginary part of the GF  $\cals G(r,r;\varepsilon)$ over the real-space coordinate $r$ as,
\begin{align}\label{eq:DoS}
n(\varepsilon)=\pm\frac{\hat j^2}{\pi}\int\td r \text{Im} \lrs{\cals G^{++}(r,r;\varepsilon)+\cals G^{--}(r,r;\varepsilon)},
\end{align}
where $\hat j^2 = 2j+1$. Numerically, the above integration was performed with a radial cutoff $R_{\max}=20$ fm and a step size of 0.1 fm. For bound states with $\Gamma/2 = 0$, a smoothing parameter $\epsilon = 10^{-6}$ MeV is introduced as the imaginary part of the probe energy $\varepsilon$ to display the DoS as the Cauchy-Lorentz distribution, instead of a sharp $\delta$-function. For resonant states, the radial integration in Eq. (\ref{eq:DoS}) undergoes a sign transition from positive to negative, when the imaginary part of the probing energy $\varepsilon$ changes from $\tIm \varepsilon>\Gamma/2$ to $\tIm \varepsilon<\Gamma/2$, with the $\pm$ sign ensuring positive-definite DoS. Practically, such sign transition is applied to identify the half-widths of resonant states. 

In contrast to bound states, one has to sample the probing energy $\varepsilon$ over the complex energy plane to identify resonant states. Specifically, by scanning the real axis, the maximum of the DoS gives an approximate energy of a resonant state. Further scanning the $\operatorname{Re} \varepsilon$ in the vicinity of the approximate energy, the half-widths $\Gamma/2$ are obtained from the sign transitions. Eventually, the energy and width of resonant state are determined by the maximum of the DoS. For additional details, please refer to Refs. \cite{SunTT2020PRC101.014321,SunTT2020CPC44.084105,ZhangY2025CPC}.

\section{Results and Discussions}\label{sec:result and discussions}

\subsection{Numerical verification}

For the verification of the RHF-GF method, the test calculation is performed for $^{120}$Sn, as compared to the methods in Refs. \cite{YangW2024APS73.062102, SunTT2016JPG43.045107, GuoJY2024PRC110.014323, CaoLg2002PRC66.024311, ZhangSS2007EPJA32.43, GuoJY2014PRC89.034307}. Figure \ref{fig:DoSSn120} displays the DoS $n_\kappa (\varepsilon)$ for the neutron orbits in $^{120}$Sn. The results are given by the RHF-GF method with the RHF Lagrangian PKO1, considering a smoothing parameter $\epsilon=10^{-6}$ MeV as the imaginary part of the probing energy $\varepsilon$. 

\begin{figure}[htbp]
  \centering
  \includegraphics[width=1.0\linewidth]{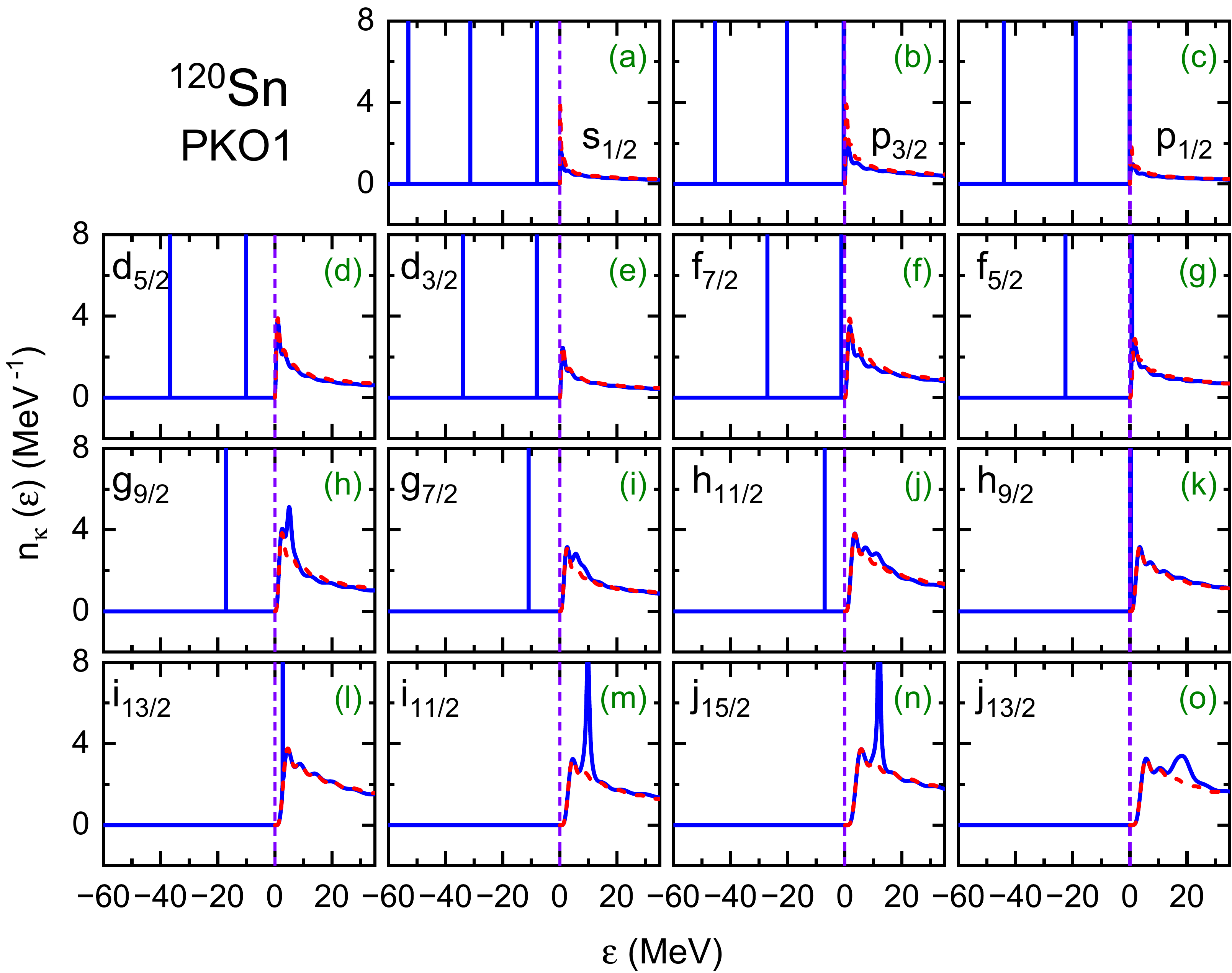}
 \caption{(Color online) Density of states $n_\kappa(\varepsilon)$ (MeV$^{-1}$) as functions of the energy $\varepsilon$ (MeV) for the neutron orbits of $^{120}$Sn. The results are calculated by the RHF-GF method with PKO1 and $R_{\rm max}$=20 fm (blue solid lines), in comparison with the result for the free-particle background obtained (red dotted lines).}\label{fig:DoSSn120}
\end{figure}

In Fig. \ref{fig:DoSSn120}, the blue solid line and red dotted line represent the DoS distributions for neutrons in the $^{120}$Sn nucleus and free particles, respectively, and the purple dashed vertical lines denoting the threshold of the continuum. The peaks below this threshold correspond to single-particle bound states, whose energies can be directly determined from the peak positions. It is checked that the bound states given by the RHF-GF method are identical to the solutions of the radial Dirac equation (\ref{eq:RHF}) with the box boundary condition.  Above the threshold lies the continuum region, which includes both resonant states and non-resonant scattering states. For the resonant states, single-particle resonance peaks can be identified by comparing the DoS of nucleons inside the nucleus (blue solid line) with that of free particles (red dotted line). Specifically, the resonant states are observed in the neutron orbits $f_{5/2}$, $h_{9/2}$, $h_{11/2}$, $g_{7/2}$, $g_{9/2}$, $i_{11/2}$, $i_{13/2}$, $j_{13/2}$, and $j_{15/2}$. For the $s$ and $p$ orbits in Figs. \ref{fig:DoSSn120}(a-c), due to the vanishing or low centrifugal barriers, it is hard to form a resonant state. 

As displayed in Fig. \ref{fig:DoSSn120}, the resonance energies and widths of these resonant states can be roughly estimated from the peak positions and full widths at half maximum of the corresponding resonance peaks. The sharp peaks in Figs. \ref{fig:DoSSn120}(g), \ref{fig:DoSSn120}(k) and \ref{fig:DoSSn120}(i) identify rather narrow resonant states in the orbits $f_{5/2} $, $h_{9/2} $ and $i_{13/2}$, respectively, which lie a bit far below the centrifugal barriers. For these narrow resonances, the barrier penetrations are suppressed, leading to long lifetimes, and their wave functions are of rather similar asymptotic behaviors as bound states. Further, the orbits $ i_{11/2} $ and $ j_{15/2} $ in Figs. \ref{fig:DoSSn120}(m) and \ref{fig:DoSSn120}(n) exhibit a typical resonance profile, which lies just below the centrifugal barrier. For these resonances, the penetration probabilities are relatively high, resulting in short lifetimes. Besides, there also exist some irregularly low peaks in Figs. \ref{fig:DoSSn120}(h-j) and \ref{fig:DoSSn120}(o), which correspond to wide resonant states in the $g$, $ h_{11/2} $ and $j_{13/2}$ orbits. These resonance energies typically exceed the centrifugal barrier, leading to extremely short lifetimes, and the asymptotic behaviors of the wave functions differ significantly from bound states. 

\begin{figure}[htbp]
  \centering
  \includegraphics[width=0.9\linewidth]{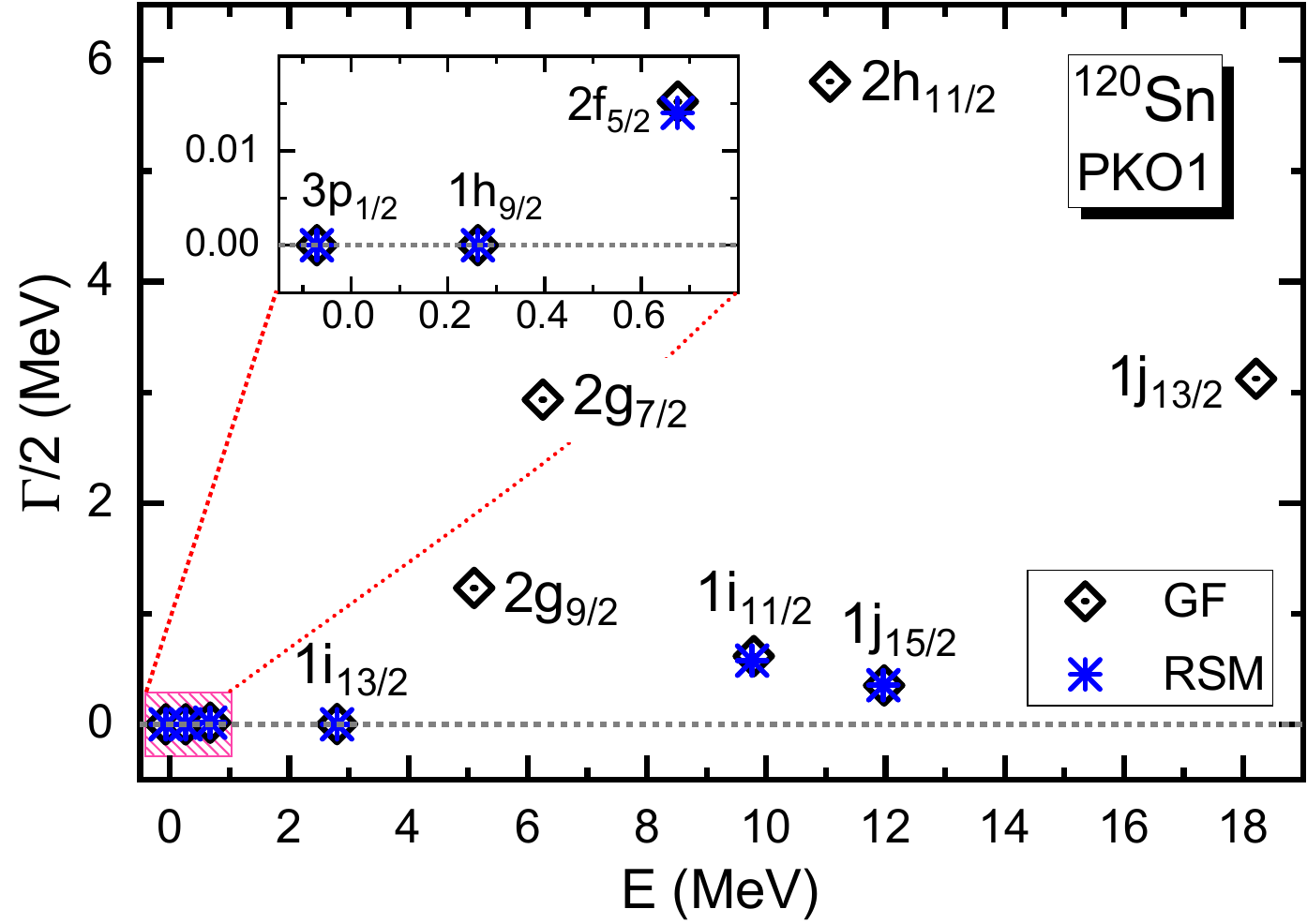}
  \caption{(Color online) Energies $E$ (MeV) and half-widths $\Gamma/2$ (MeV) of the neutron resonant states and weakly bound one $3p_{1/2}$ in $^{120}$Sn. The results are obtained by the RHF theory with the Green's function (GF) method, as compared to those with the real stabilization method (RSM) \cite{YangW2024APS73.062102}.}\label{fig:Sn120Reslev}
\end{figure}

For rather wide resonances, the resonance energy and width estimated from Fig. \ref{fig:DoSSn120} are subject to relatively large uncertainties. The high-precision energy and half-width of the resonant states can be further determined by using the maximum and the positive-to-negative sign transition of the radial integration in Eq. (\ref{eq:DoS}), following a similar approach as that illustrated in Figs. 2 and 3 of Ref. \cite{SunTT2020CPC44.084105}, with the corresponding results presented in Fig. \ref{fig:Sn120Reslev}. The results given by the RHF-RSM method \cite{YangW2024APS73.062102} are shown for comparison, and the inset is an enlarged view of the bottom-left shadowed region for the low-lying resonant states. Given the oscillations at $r \to \infty$, the principal quantum numbers deducing from the nodes of wave functions no longer hold for resonant states. Here an energy-ordered number, inheriting from bound states, is used to denote resonant states. 

\begin{table*}[htbp]
\caption{Energies $E$ (MeV) and widths $\Gamma$ (MeV) of the proton resonant states in $^{120}$Sn. The results are obtained by the RHF theory with the Green's function (GF) method, in comparison with those obtained by the RMF theory with the GF method \cite{SunTT2016JPG43.045107}, CMR method \cite{GuoJY2024PRC110.014323}, $S$-matrix method \cite{CaoLg2002PRC66.024311}, ACCC method \cite{ZhangSS2007EPJA32.43}, and CSM \cite{GuoJY2014PRC89.034307}. }\label{tab:Sn120Reslev}
\setlength{\tabcolsep}{0.8em}\renewcommand{\arraystretch}{1.4}
\begin{tabular}{c|rr|rr|rr|rr|rr|rr} \hline\hline 
\multirow{2}{*}{$nl_j$} &\multicolumn{2}{c|}{ RHF with GF} & \multicolumn{2}{c|}{GF \cite{SunTT2016JPG43.045107}} & \multicolumn{2}{c|}{CMR \cite{GuoJY2024PRC110.014323}} & \multicolumn{2}{c|}{$S$-matrix \cite{CaoLg2002PRC66.024311}} & \multicolumn{2}{c|}{ACCC \cite{ZhangSS2007EPJA32.43}}  & \multicolumn{2}{c}{CSM \cite{GuoJY2014PRC89.034307}}    \\
& $E$~~~~   & $\Gamma$~~~~   & $E$~~~~& $\Gamma$~~~~& $E$~~~~& $\Gamma$~~~~& $E$~~~~& $\Gamma$~~~~& $E$~~~~& $\Gamma$~~~~&$E$~~~~& $\Gamma$~~~~\\ \hline
$3p_{ 3/2}$ &  6.801 & 0.617 &  7.265 & 0.965     &  7.567 & 1.291  &  7.513 & 0.924       & 7.320  & 0.820       & 7.305  & 0.911       \\
$3p_{ 1/2}$ &  7.168 & 0.933 &  7.667 & 1.233     &  8.166 & 2.052  &  8.085 & 1.344       & 7.690  & 1.130       & 7.663  & 1.222       \\
$2f_{ 7/2}$ &  5.648 & 0.017 &  6.205 & 0.037     &  6.210 & 0.042  &  6.210 & 0.043       & 6.220  & 0.073       & 6.207  & 0.048       \\
$2f_{ 5/2}$ &  7.380 & 0.175 &  7.909 & 0.365     &  7.917 & 0.294  &  7.934 & 0.307       & 7.970  & 0.300       & 7.919  & 0.282       \\
$1h_{ 9/2}$ &  6.842 & 0.002 &  7.134 & 0.002     &  7.133 & 0.003  &  7.132 & 0.003       & 7.130  & 0.017       & 7.135  & 0.003       \\
$1i_{13/2}$ &  9.767 & 0.008 & 10.110 & 0.014     & 10.110 & 0.012  & 10.110 & 0.012       & $-$~~~~& $-$~~~~     & $-$~~~~& $-$~~~~     \\
$1i_{11/2}$ & 16.784 & 0.895 & 16.934 & 1.092     & 16.889 & 0.946  & 16.960 & 0.999       & $-$~~~~& $-$~~~~     & $-$~~~~& $-$~~~~     \\
$1j_{15/2}$ & 19.325 & 0.621 & 19.779 & 0.730     & $-$~~~~& $-$~~~ & $-$~~~~& $-$~~~~     & $-$~~~~& $-$~~~~     & $-$~~~~& $-$~~~~     \\  
$2g_{ 9/2}$ & 13.117 & 2.483 &  $-$~~~& $-$~~~~   & 13.510 & 3.208  & 14.772 & 3.502       & $-$~~~~& $-$~~~~     & $-$~~~~& $-$~~~~     \\
\hline\hline
\end{tabular}
\end{table*}

As illustrated in Fig. \ref{fig:Sn120Reslev}, the resonant states with extremely short lifetimes, such as $2g_{7/2}$, $2g_{9/2}$, $2h_{11/2}$ and $1j_{13/2}$, are not captured by the RHF-RSM method, which works well for the resonant states with $\Gamma/2<1$ MeV, being identical with those obtained by the RHF-GF method. For the spin partners, the splittings of the $1i$ and $1j$ resonant states remain significant, but much less for the $2g$ ones. Moreover, the resonant states with $j_<=l-1/2$ lie higher than those with $j_>=l+1/2$. This suggests that, similar to bound states, the spin-orbit coupling also plays a significant role in resonant states, while the spin-orbit splitting could be essential to their broadening. For instance, the $2g$ resonant states with $\Gamma/2>1$ exhibit fairly small splitting, in contrast to the $1i$ and $1j$ ones. 

In order to provide an implemented illustration, Table~\ref{tab:Sn120Reslev} lists the energies $E$ and widths $\Gamma$ of the proton resonant states of $^{120}$Sn obtained by the RHF-GF method, in comparison with those given by the RMF theory with the GF method \cite{SunTT2016JPG43.045107}, CMR method \cite{GuoJY2024PRC110.014323}, $S$-matrix method \cite{CaoLg2002PRC66.024311}, ACCC method \cite{ZhangSS2007EPJA32.43}, and CSM \cite{GuoJY2014PRC89.034307}. Due to the improvement of the strategy in searching resonant states \cite{SunTT2020CPC44.084105}, an additional resonant state $2g_{9/2}$ is obtained by the RHF-GF method, similar to the CMR and $S$-matrix methods. In fact, using the RMF-based GF method, we also obtain the energy and width for the resonant state $2g_{9/2}$, namely $E=13.449$ MeV and $\Gamma=3.140$ MeV, which were not provided in Ref. \cite{SunTT2016JPG43.045107}. On the other hand, the energies and widths obtained by the RHF-GF method are systematically smaller than those given by the RMF-based methods. This may arise from the Fock terms, which change the modeling of nuclear binding from the RMF models as pointed out in Ref. \cite{Geng2019PRC100.051301R}.

\subsection{Coulomb exchange effects on proton resonances}

In order to understand the Coulomb exchange effects, the even isotones of $N=82$ from $Z=44$ to $54$ are taken as examples. Starting from the converged RHF calculation, the RHF-GF calculations were conducted with and without the Coulomb exchange terms, giving the proton resonance energies and widths as $(E, \Gamma)$ and $(E^0,\Gamma^0)$, respectively. The differences $\Delta E = E^0 - E$ and $\Delta \Gamma = \Gamma^0 - \Gamma$ quantify the Coulomb exchange reductions in the resonance energies and widths, respectively. 

\begin{figure}[htbp]
  \centering
  \includegraphics[width=0.92\linewidth]{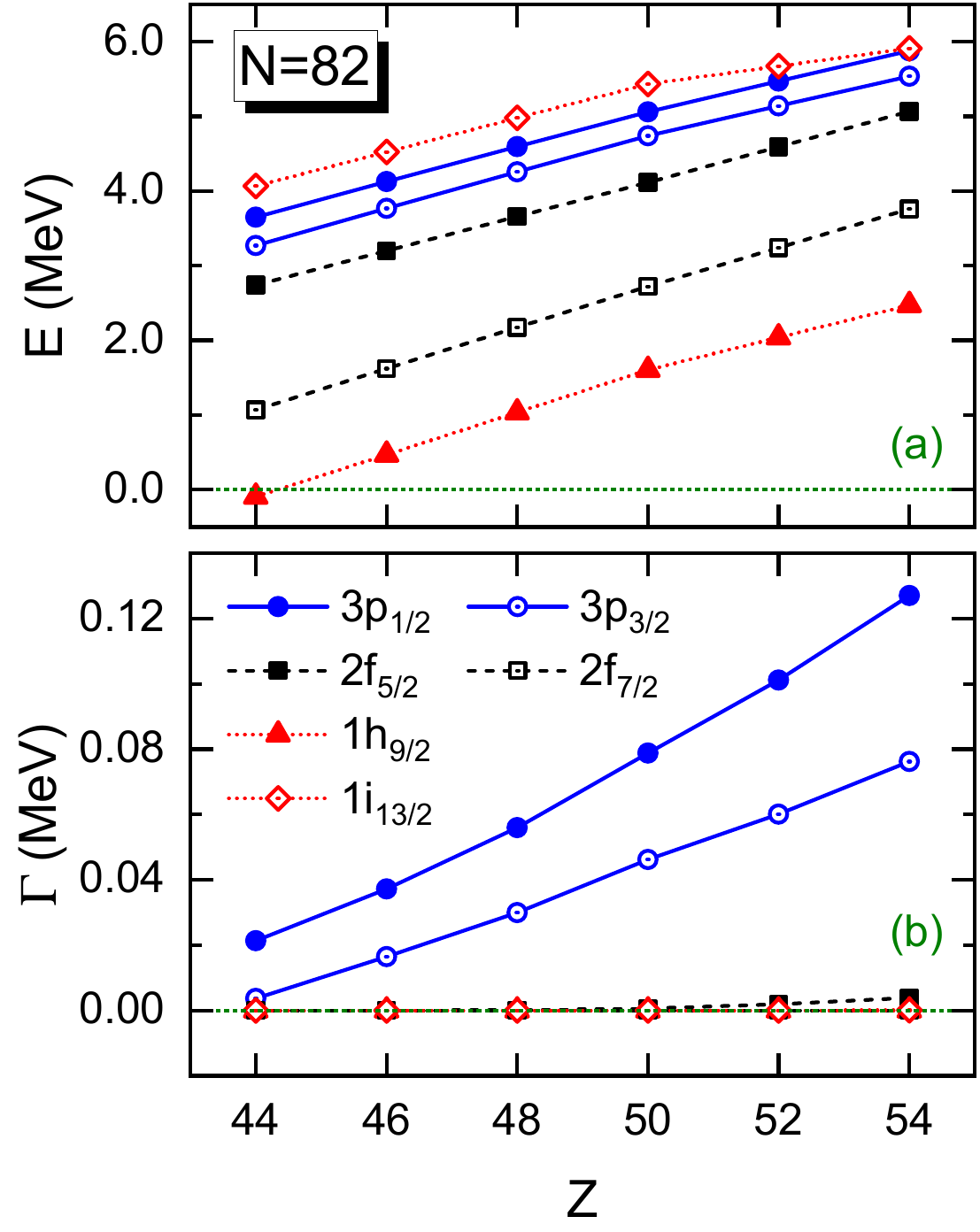}
  \caption{(Color online) Plots (a) and (b) display the energies $E$ (MeV) and widths $\Gamma$ (MeV) of the proton resonant states for the $N=82$ isotones, respectively. The results are calculated by the RHF theory with the GF method, in which the Coulomb exchange terms are treated exactly. }\label{fig:82N.Er_Gam}
\end{figure}

Figure \ref{fig:82N.Er_Gam} displays the isotonic evolution of the proton resonance energies and widths of the $N=82$ isotones, given by the complete RHF-GF calculations with PKO1. Due to gradually enhanced Coulomb repulsions, the resonance energies in Fig. \ref{fig:82N.Er_Gam}(a) illustrate a monotonous and nearly parallel increasing with respect to the proton number. As an exception, the increasing of the resonance energies of $1i_{13/2}$ is slightly reduced when $Z > 50$, which is also observed in the RMF-GF calculations and the RHF-GF ones excluding the Coulomb exchange terms. Moreover, as shown in Fig. \ref{fig:82N.Er_Gam}(b), the widths of the narrow resonances $2f_{7/2}$, $2f_{5/2}$, $1h_{9/2}$ and $1i_{13/2}$ remain almost unchanged, due to their relatively high centrifugal barriers. In contrast, due to fairly high energies and low centrifugal barriers, the resonance widths of the $3p$ states visually increase with respect to the proton number. In fact, similar systematics are also obtained by the RMF-CSM \cite{Niu2013PRC87.037301,NiuZM2016NST27.122} and RMF-GF \cite{SunTT2016JPG43.045107} calculations. This seems to indicate that the Coulomb exchange terms do not exert substantial effects on the isotonic behaviors of proton resonance energies and widths.

\begin{figure}[htbp]
  \centering
  \includegraphics[width=.95\linewidth]{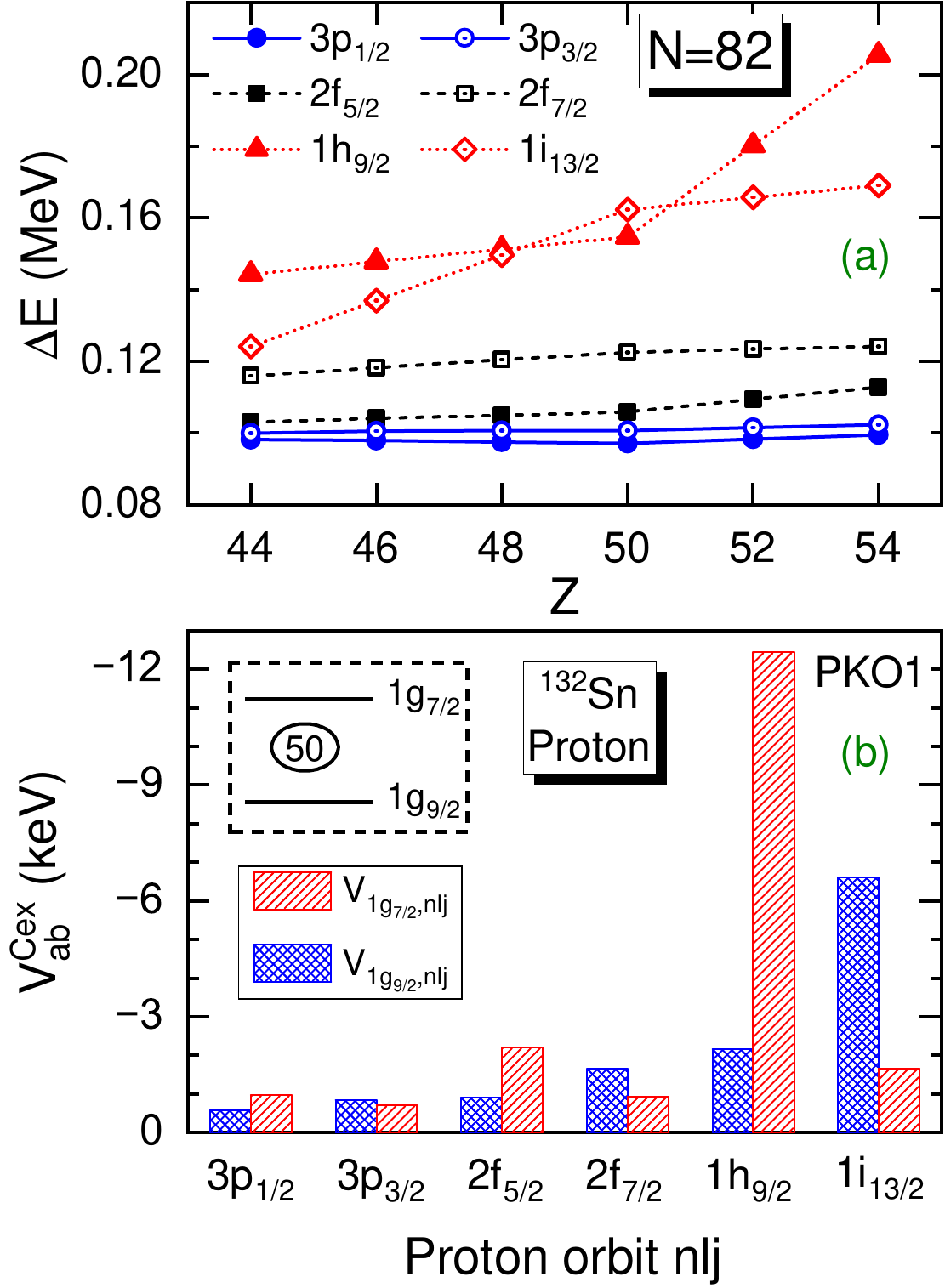}
  \caption{(Color online) (a) Coulomb exchange reductions $\Delta E$ (MeV) in resonance energies for proton resonant states of the $N=82$ isotones; (b) Interaction matrix elements $V_{ab}^{\text{Cex}}$ (keV) contributed by the Coulomb exchange terms between proton bound $1g$ states and resonant states. The results are given by the RHF-GF method with PKO1.}\label{fig:82N.Er_Gam_diff}
\end{figure}

To provide further insight, Fig. \ref{fig:82N.Er_Gam_diff}(a) displays the Coulomb exchange reductions of the proton resonance energies $\Delta E$ (MeV), which range from 0.09 to 0.21 MeV. It is worth noting that, using phenomenological Coulomb exchange terms, the resonance energy reductions can be as large as about 0.5 MeV \cite{GuoJY2014PRC89.034307,NiuZM2016NST27.122}. Different from the $2f$ and $3p$ resonances, the values of $\Delta E$ for the high-$l$ resonances $1h_{9/2}$ and $1i_{13/2}$ increase progressively with proton number, which exhibit pronounced but opposite kinks at $Z=50$, suggesting distinct shell effects. For better understanding, applying the non-local densities (\ref{eq:GF_non-local_density}), the interaction matrix elements contributed by the Coulomb exchange terms, namely $V_{ab}^{\text{Cex}}$ (keV), are extracted from the RHF-GF calculations.  Taking $^{132}$Sn as a candidate, Fig. \ref{fig:82N.Er_Gam_diff}(b) shows the $V_{ab}^{\text{Cex}}$ values for the couplings between the bound states $1g$ and resonances $3p$, $2f$, $1h_{9/2}$ and $1i_{13/2}$. As deduced from the inset in Fig. \ref{fig:82N.Er_Gam_diff}(b), the proton shell closure $Z=50$ is given by the spin-orbit splitting of the $1g$ orbits.

As illustrated in Fig. \ref{fig:82N.Er_Gam_diff}(b), via the Coulomb exchange terms, the resonant state $1h_{9/2}$ shows much stronger coupling with the $1g_{7/2}$ state than with the $1g_{9/2}$ one. This explains well the kink at $Z=50$ in Fig. \ref{fig:82N.Er_Gam_diff}(a), as combined with the fact that the protons populate orderly the $1g_{9/2}$ and $1g_{7/2}$ states. Similarly, it is also understandable for the opposite kink in the $\Delta E$ values for the $1i_{13/2}$ state. In fact, for the discussed resonant states here, there exists certain correspondence between the order numbers of the states and the nodes of the wave functions. Given the nodal differences, it is easy to understand why rather small $V_{ab}^{\text{Cex}}$ values are obtained for the couplings between the bound states $1g$ and the resonances $3p$ and $2f$, as illustrated in Fig. \ref{fig:82N.Er_Gam_diff}(b). For the differences given by the couplings of the spin partners $1g_{9/2}$ and $1g_{7/2}$ with the resonance $1h_{9/2}$, as well as with $1i_{13/2}$, this can be attributed to the spin-orbit effects carried by the Coulomb exchange terms. Meanwhile, it shall be stressed that, the microscopic shell effects observed in Fig. \ref{fig:82N.Er_Gam_diff}(a) were not found in the RMF-CSM calculations using phenomenological formula \cite{NiuZM2016NST27.122}, which underscores the necessity of exact treatment of Coulomb exchange terms.

\begin{figure}[htbp]
  \centering
 \includegraphics[width=1.0\linewidth]{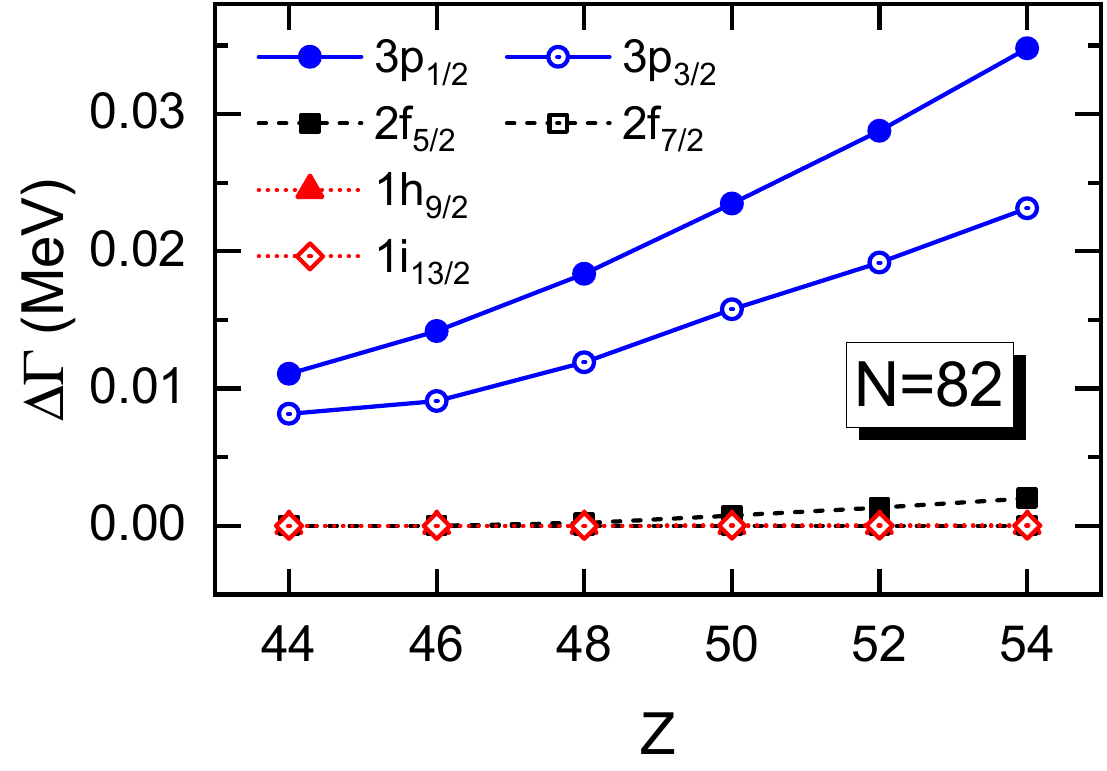}
 \caption{(Color online) Coulomb exchange reductions $\Delta\Gamma$ (MeV) in resonance widths for the proton resonant states of the $N=82$ isotones. The results are given by the RHF-GF method with PKO1.}\label{fig:DENabc}
\end{figure}

Furthermore, Fig. \ref{fig:DENabc} show the Coulomb exchange reductions of resonance widths $\Delta\Gamma$ (MeV) for the proton resonant states of the $N=82$ isotones. Compared to the narrow resonances $2f$, $1h_{9/2}$ and $1i_{13/2}$, it is obvious that the Coulomb exchange terms exert much stronger reductions in the widths for the resonance $3p$, which increase progressively with proton number. This indicates that the impact on the widths primarily depends on the intrinsic width of the resonance itself. This is different from the reductions of the energies, which are primarily determined by the interactions as illustrated in Fig. \ref{fig:82N.Er_Gam_diff} (b).  However, if looking at the resonance lifetime, which is inversely proportional to the width, namely $\tau \simeq \hbar/\Gamma$ \cite{book.Kukulin1989}, the effects of the Coulomb exchange terms can be visible for narrow resonances, despite the tiny reductions of the widths. Taking $^{136}$Xe as an example, neglecting the Coulomb exchange terms reduce the lifetime of the resonance $1i_{13/2}$ by $\sim$37.8\%, namely from $3.7 \times 10^{-2}$ fs to $2.3 \times 10^{-2}$ fs, in contrast to the $\sim$11.5\% reduction of the lifetime of $3p_{1/2}$, which decreases from $5.2 \times 10^{-6}$ fs to $4.6 \times 10^{-6}$ fs.

Intuitively, the Coulomb exchange terms reduce the Coulomb repulsion experienced by protons. With the inclusion of the Coulomb exchange terms, the proton resonance energies become smaller, and the resonance widths are supposed to reduce consistently. Meanwhile, the Coulomb potential barrier becomes lower and thinner to facilitate the barrier penetration, an effect enlarging the resonance widths. Thus, proton resonance energies are reduced by the Coulomb exchange terms, whereas the impact on the resonance widths stems from the competition of these two effects. Our results illustrate that, with the presence of the Coulomb exchange terms, both the proton resonance energies and widths are reduced, despite the lower and thinner Coulomb barrier.

\section{Summary}\label{sec:summary}

In this work, the GF  method has been applied to the RHF theory to provide a unified description of both bound and resonant states, where resonance energies and widths are accurately extracted from the DoS. As a benchmark, neutron and proton resonances of $^{120}$Sn are calculated using the RHF-GF method, yielding consistent results with those from the RHF-RSM and RMF based methods. 

Furthermore, the Coulomb exchange effects on proton resonances are investigated by taking the even $N=82$ isotones as test cases. It is illustrated that the exact Coulomb exchange terms incorporated in the RHF-GF method reduce the  proton resonance energies by approximately $0.09\sim0.21$ MeV, significantly less than those given by the phenomenological Coulomb exchange terms. Moreover, except for rather narrow resonance, visible width reductions are also observed, being substantially smaller than those resulting from the phenomenological approach. Particularly, the energy reductions exhibit distinct shell effects, highlighting the significance of a microscopic and exact treatment of the Coulomb exchange terms. This also indicates that the RHF-GF method, by incorporating \add{the} Fock terms, lays the foundation for exploring single-particle resonances, further for the reliable modeling of nucleon emission, capture, and scattering processes. 

\begin{acknowledgments}
  This work was partly supported by the National Natural Science Foundation of China under Grant No. 12275111, the Fundamental Research Funds for the Central Universities lzujbky-2023-stlt01, and the Supercomputing Center of Lanzhou University. 
\end{acknowledgments}
%
%

\end{document}